\documentstyle[floats,aps,epsf]{revtex}

\title{ Clustering in stable and unstable nuclei in $p$-shell and $sd$-shell 
regions}

\author{Y. Kanada-En'yo}
\address{Institute of Particle and Nuclear Studies,\\
High Energy Accelerator Research Organization, \\
Oho 1-1, Tsukuba-shi 305-0801, Japan}
\author{M. Kimura}\address{Institute of Physical and Chemical Research(RIKEN),
 Saitama 351-0198, Japan}
\author{H. Horiuchi}\address{Department of Physics, Kyoto University,\\
Kitashirakawa-Oiwake, Sakyo-ku, 
Kyoto 606-01,
Japan}

\begin{document}

\maketitle

\begin{abstract}
According to microscopic calculations with 
antisymmetrized molecular dynamics, we studied cluster features 
in stable and unstable nuclei. 
A variety of structure was found in stable and unstable nuclei in the
$p$-shell and $sd$-shell regions.
The structure of excited states of $^{12}$Be was
investigated, while in $sd$-shell nuclei we focused on molecular states
and deformed states.
The deformed states in $^{28}$Si and
$^{40}$Ca were discussed in connection with the high-lying molecular states. 
Appealing molecular states in $^{36}$Ar
and $^{24}$Mg were suggested. 
The results signified that both clustering of nucleons and 
mean-field formation are essential features
in $sd$-shell nuclei as well as $p$-shell nuclei.
\end{abstract}

\section{Introduction}
Clustering is one of the essential features in 
nuclear dynamics, as already seen in light stable nuclei, where
various cluster structure has been known even 
in the low-energy region.
Also in the physics of 
unstable nuclei, the cluster feature is one of the center subjects.
In fact, the recent experimental and theoretical studies of neutron-rich Be 
isotopes \cite{TANIHATA,FREER,SAITO,SEYA,OERTZEN,ARAI,DOTE,ENYOg,ITAGAKI,OGAWA,ENYObe14,ENYObe12,ITO,DESCOUVEMONT} revealed that cluster states exist also 
in light unstable nuclei.
On the other hand, needless to say, the mean-field nature is the 
other essential aspect, which become more important in heavier nuclei. 
These facts imply that a variety of structure appears in stable and
unstable nuclei.
In the $p$-shell region, both the cluster and mean-field aspects coexist in a
nucleus such as $^{12}$C \cite{ENYOe}. 
However, it is an open question whether or not the cluster features 
survive and are still essential in heavier nuclei.
Concerning cluster aspect in stable $sd$-shell nuclei, remarkable phenomena
 are the molecular resonances, which have been observed
 in highly excited states
of such nuclei as $^{24}$Mg, $^{28}$Si and $^{32}$S
\cite{JAMES,OHKUBO-O16}.
When we ascertain the possible coexistence of cluster and mean-field natures 
in $sd$-shell or heavier nuclei, one of the key issues is to connect
the molecular resonances with the low-lying deformed states. 

Our aim is to make a systematic study covering such a variety of structure
in the wide mass number region, while focusing on the cluster features.
The important thing in theoretical models is 
describing both the cluster structure and mean-field-type structure.
We apply microscopic methods based on antisymmetrized molecular 
dynamics(AMD)\cite{ENYOsup,AMDrev}, 
which meet the requirement because of the flexibility of the
AMD wave functions. In this paper, the excited states of $^{12}$Be, 
the shape coexistence and molecular resonances in $^{28}$Si and $^{40}$Ca
are studied. In $^{36}$Ar and $^{24}$Mg, possible molecular states 
in highly excited states are suggested. 

\section{Formulation}
The formulation of AMD for nuclear structure study of the ground 
and excited states is explained in Refs.\cite{ENYOsup,AMDrev}.
The wave function of a nucleus is written by AMD wave functions,
$\Phi=c \Phi_{AMD} +c' \Phi '_{AMD} + \cdots$.
An AMD wave function 
is a Slater determinant of Gaussian wave packets;
\begin{eqnarray}
&\Phi_{AMD}({\bf Z})=\frac{1}{\sqrt{A!}}
{\cal A}\{\varphi_1,\varphi_2,\cdots,\varphi_A\},\\
&\varphi_i=\phi_{{\bf X}_i}\chi_{\xi_i}\tau_i, {\rm where\quad}
\phi_{{\bf X}_i}({\bf r}_j) \propto
\exp\left 
[-\nu\biggl
({\bf r}_j-\frac{{\bf X}_{i}}{\sqrt{\nu}}\biggr)^2
\right] {\rm and \quad } \chi_{\xi_i}=\left(\begin{array}{l}
{1\over 2}+\xi_{i}\\
{1\over 2}-\xi_{i}
\end{array}\right).
\end{eqnarray}
$\chi_{\xi_i}$ is the intrinsic spin function and 
$\tau_i$ is the iso-spin function. 
An AMD wave function is given by a set of complex 
variational parameters
${\bf Z}=\lbrace {\bf X}_1, {\bf X}_2, \cdots, {\bf X}_A, \xi_1, 
\cdots, \xi_A \rbrace$, where ${\bf X}_i$ is the center of Gaussian and
${\xi_i}$ defines the orientation of the intrinsic spin 
for the $i$-th single particle wave function.
The model space of AMD wave function covers 
various cluster structure and also shell-model-like states.
We perform energy variation by the frictional cooling method with respect to
the complex variational parameters to obtain an energy minimum state 
in the model space. 
For the excited states of $^{12}$Be isotopes, we performed 
energy variation after spin-parity projection(VAP). 
In case of $sd$-shell nuclei, we applied a generator coordinate method
in the AMD framework(AMD+GCM).
Harmonic-oscillator quanta 
of the system, $\langle \sum^A_{i=1}
{\bf p}_i^2/2m\hbar\omega+m\omega{\bf r}^2_i/2\hbar\rangle$,
is adopted as the generator coordinate in the present paper.

\section{Excited states of $^{12}$Be}
\begin{figure}[t]
\epsfxsize=0.5\textwidth
\centerline{\epsffile{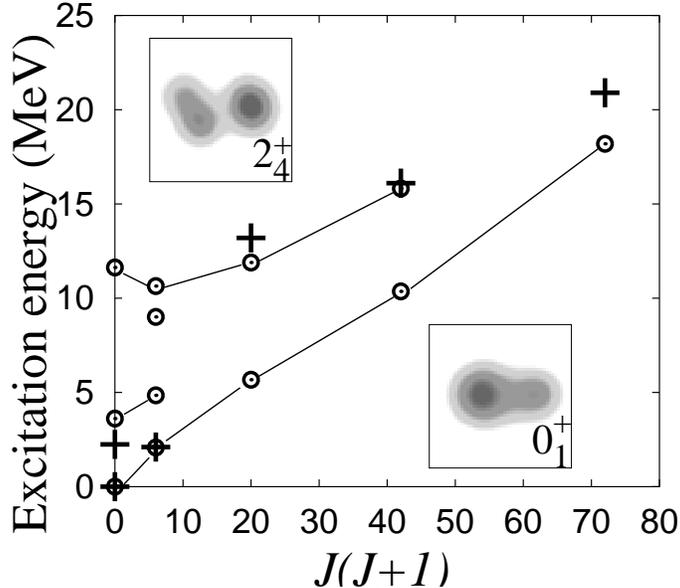}}
\caption{
Energy-spin systematics for positive-parity states in $^{12}$Be.
The excitation energies of the natural spin states are plotted as 
functions of $J(J+1)$. The circle symbols are the 
theoretical results, while the cross symbols are 
the experimental data of the spin-parity assigned states 
taken from the Table of Isotopes and 
Refs. \protect{\cite{FREER,SHIMOURA}}.}
 \label{fig:be12rot}
\end{figure}

$^{12}$Be is an interesting nucleus where the vanishing of magic number $N=8$
occurs. Recently, many excited states of $^{12}$Be have been experimentally
measured\cite{TANIHATA,FREER,SAITO,SHIMOURA}.
We applied the VAP method to the excited states of Be isotopes by using 
effective interactions, MV1 force(case 1, $m=0.65$)+G3RS-type LS force
($u_1=-u_2=3700$ MeV)+Coulomb.
In Fig.\ref{fig:be12rot}, the postivie-parity rotational bands
are diplayed in comparison with the experimental data of the 
spin-assigned positive-parity states.
The highly excited states 
measured by He-He breakup reactions \cite{FREER} are candidates of 
molecular-like states. 
Theoretical results obtained by the VAP calculation 
well fit to the experimental data.
We found three rotational bands 
$K^\pi=0^+_1, 0^+_2$ and $0^+_3$ in the positive-parity states.
The ground band consists of the intruder states ($2\hbar\omega$
excited configurations), which are
well-deformed states with 2$\alpha$ core in the surrounding neutrons.
On the other hand, the normal neutron-shell-closed states belong to the
second band. It means that the vanishing of neutron magic number $N=8$
occurs in $^{12}$Be.
As a result of the inversion between the cluster states 
and the normal $p$-shell closed states, the calculation consistently agrees
with the data of the $\beta$ decay strength from the ground state into $^{12}$B
\cite{ENYObe12}. 
The other interesting point in the results is the prediction of the
$^6$He+$^6$He molecule-like structure in the third $0^+$ band.
The experimentally measured $4^+$ and $6^+$ states are the candidates of 
these $^6$He+$^6$He states in the $K^\pi=0^+_3$ band. 
Thus we found a variety of structures in the excited states of $^{12}$Be.
We make a point of {\it the coexistence of cluster and mean-field natures}
in this light neutron-rich nucleus. In the ground states, 
the valence neutrons move
around the 2$\alpha$ core. In other words, they show the 
deformed-mean-field behaviour, which causes an enhancement of 
the 2$\alpha$ clustering. On the other hand, the shell-model-like states belong
to the $K^\pi=0^+_2$ band, while the developed $^6$He-$^6$He molecular
structure was found in the $K^\pi=0^+_3$ band.

\section{$sd$-shell nuclei}
In the study of cluster features in $sd$-shell nuclei,
the subject of molecular resonances is one of the key problems.
Molecular resonances have been experimentally measured in such $Z=N$
nuclei as $^{24}$Mg, $^{28}$Si and $^{32}$S.
It is a long standing issue what is the connection between the 
molecular resonances in the high energy region and the low-lying deformed
states. In present paper, we applied AMD+GCM method for the study of 
$sd$-shell nuclei.

\begin{figure}[h]
\epsfxsize=0.75\textwidth
\centerline{\epsffile{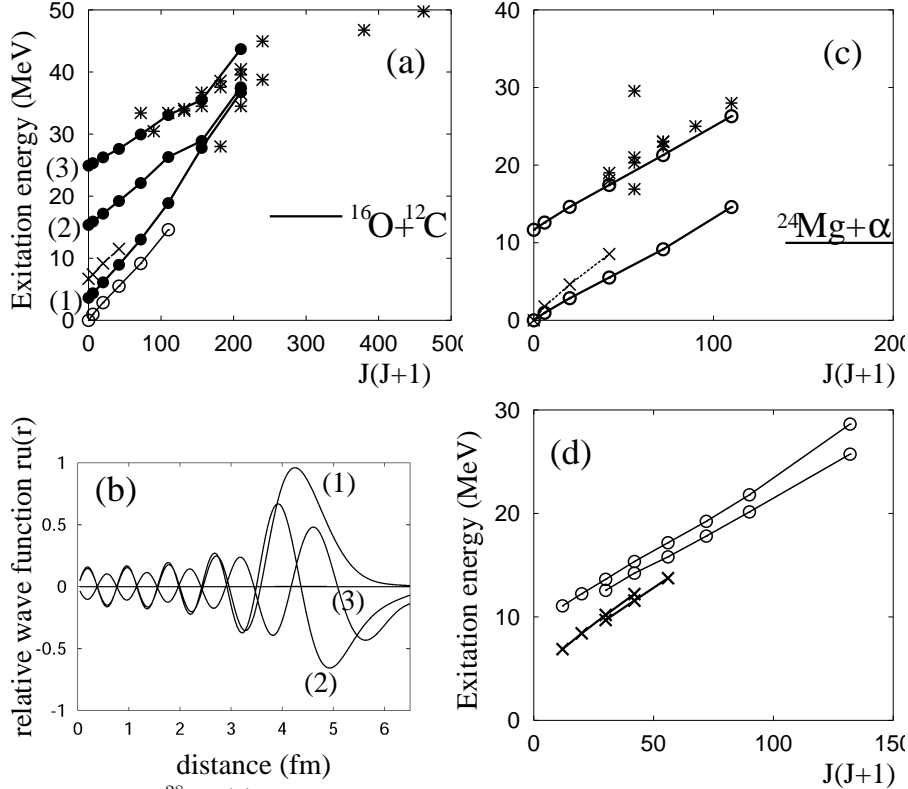}}
\caption{\label{fig:si28rot} 
Rotational bands of $^{28}$Si. 
(a) Calculated positive-parity bands dominated by O+C cluster 
components(filled circles) and the ground band(open circles).
Experimental data of the prolate band 
(the O-C molecular resonances) are demonstrated by symbols +($\ast$).
(c) Calculations of the ground band and the positive-parity
$\alpha$-cluster band(circles).
Cross points are the experimental ground band.
Symbols($\ast$) denote the experimental data observed in 
$\alpha$-transfer reactions.
(d) Low-lying negative-parity $K^\pi=3^-$ and $K^\pi=5^-$ bands. 
(b) The O-C inter-cluster wave function
in the band-head states of three rotational bands 
displayed in figure(a). 
}
 \label{fig:si28}
\end{figure}

The first subject is $^{28}$Si, where we have already known various
experimental suggestions on the shape coexistence and the molecular
resonances. In the low-energy region, the oblate ground band and 
a prolate excited band are known, while exotic shapes in the negative 
parity states are expected from the $\gamma$-ray transition strengths
\cite{GLATZ}.
In the highly excited states, 
$^{16}$O-$^{12}$C molecular resonances were experimentally observed
\cite{JAMES}.
Moreover, $\alpha$-cluster states are suggested\cite{ARTEMOV}. 
There were many theoretical efforts by mean-field approaches and cluster
approaches\cite{BAYE,BAUHOFF,KATO}, but one did not yet catch 
a total understanding of 
overall these phenomena. 
We studied the structure of the ground, excited states of $^{28}$Si
and investigated the molecular resonant behaviour
with AMD+GCM method(fixed intrinsic spins version).
First, we obtain energy curves as a function
of the generator coordinate(GC)
$\langle \sum^A_{i=1}
{\bf p}_i^2/2m\hbar\omega+m\omega{\bf r}^2_i/2\hbar\rangle$, 
harmonic oscillator
quanta.
In case of $^{28}$Si, two local minimum states are found in each value of 
GC. The lower one corresponds to the oblate deformed state and the 
higher one is the prolate deformation. In the small GC region, these states
compose the oblate ground band and the prolate excited band. 
With the increase of the GC value, the oblate deformation varies into
$\alpha$-cluster($\alpha+^{24}$Mg) states, while from the prolate deformation
the $^{16}$O+$^{12}$C clustering enlarges. After the superposition of the 
obtained basis, O-C molecular states and $\alpha$-cluster states appear from
the prolate band and the oblate ground band, respectively.
In the final results, we chose 23 AMD wave functions as base states in the
diagonalization. By analysing the amplitudes of the base states 
in the final states, we can classify many of the excited states into two
groups. The first group principally consists of the prolate states, while the 
second group is a family based on the oblate states.

In the positive-parity states of the former group(prolately deformed states),
we found three $K^\pi=0^+$ bands dominated by O+C cluster states.
We demonstrate in Fig.\ref{fig:si28rot}(a) 
the excitation energy of the three O+C cluster 
bands and the oblate ground band. The lowest band of the three is consistent 
with the normal prolately deformed band in the experimental data.
In order to discuss the molecular resonant behaviour of the higher bands,
we give in Fig.\ref{fig:si28rot}(b) the extracted inter-cluster wave functions 
between the 
$O_{g.s.}$ and $C_{g.s.}$($SU_3$ limit g.s.) clusters. It is found that the 
lowest band has the minimum node number $n_0=8$ in Pauli allowed states, 
while the second and third bands are the higher nodal
O-C molecular states with larger node numbers of the inter-cluster 
wave functions as $n_0=9$ and $n_0=10$,
respectively. In the energy region near
the third bands, there exist many experimentally observed O-C molecular 
resonances. 
It is interesting that the low-lying prolate band of $^{28}$Si 
is regarded as the lowest O+C cluster state and the higher nodal
molecular bands appear due to the excitation of the inter-cluster motion.
In fact, the molecular resonances in $^{28}$si 
were often discussed in the connection 
with the low-lying prolate deformation in the studies 
with two-body cluster models
\cite{BAYE,KATO}. Here we remark the importance of the present work 
that we did confirm the appearance of O-C molecular states 
without assuming the existence of any clusters. 

In the family of the oblate states, we obtain an excited band consisting of 
$\alpha$-cluster states in addition 
to the oblate ground band(Fig.\ref{fig:si28rot}(c)). 
The $\alpha$-cluster band is
about 10 MeV higher than the ground band in the present results.
There are experimental levels observed in $\alpha$-transfer reactions
\cite{ARTEMOV}
in the corresponding energy region. In the negative parity states,
the calculated results indicate that exotic shapes compose $K^\pi=3^-$
and $K^\pi=5^-$ bands(Fig.\ref{fig:si28rot}(d)). 
For example, $K^\pi=5^-$ band is described by 
a pentagon shape given by 7$\alpha$-like structure. The results agree
to the experimental data of the energy levels and the in-band $E2$ 
transition strengths of the $K^\pi=3^-$ and $K^\pi=5^-$ bands.

The second subject is $^{40}$Ca, where, 
the rotational band of the 
superdeformation has been recently observed in the 
experimental data of $\gamma$-ray measurements \cite{IDEGUCHI}.
It is very surprising that many-particle many-hole states
lie in such a low-energy region as the excitation energy of
the band-head state of this superdeformation is only 5.2 MeV.
The low-lying excited states of $^{40}$Ca were studied by the 
AMD+GCM method in Ref.\cite{ENYOca40}, where 
many low-lying rotational bands were found in the results.
By analysing the single-particle orbits of the positive-parity states, 
it was found that 
the ground band, the normal deformation and the superdeformation
are dominated by $0p$-$0h$, $4p$-$4h$ and $8p$-$8h$ 
configurations, respectively. 
It is consistent with the analysis based on shell-model 
calculations\cite{IDEGUCHI}. What is exciting in our results
is that superdeformation has a parity-asymmetric shape
which is composed by the $^{28}$Si+$^{12}$C-like cluster core.
It is unique that, in the
superdeformation, the parity-asymmetric shape like a pear arises 
keeping the $8p$-$8h$ configurations dominant. With respect to
the parity-asymmetric shape, we predicted the parity-doublet states
in the negative parity states, which contain the mixing of $7h$-$7p$ 
and $5h$-$5p$ configurations \cite{ENYOca40}.
Moreover, it is naturally expected that the $^{28}$Si-$^{12}$C molecular 
states may arise from the superdeformation. From this point of view, 
we studied the highly excited states of $^{40}$Ca in the present paper. 
As we increase the generator coordinate
(harmonic oscillator of the system) from the superdeformed state, 
the $^{28}$Si+$^{12}$C clustering enhances. After the diagonalization,
we found the higher rotational bands with the remarkable 
$^{28}$Si-$^{12}$C molecular states. 
Namely, the $^{28}$Si-$^{12}$C molecular states appear 
due to the excitation of inter-cluster motion based on the superdeformation.
In the other words, the low-lying superdeformation is regarded as the lower
member of the $^{28}$Si-$^{12}$C molecular states. 
This is a similar situation to the $^{28}$Si system, where
the low-lying deformed states are connected with the O-C molecular 
resonances.
In the present work, the energy intervals between the superdeformed and 
the second bands, the second and the third bands are predicted to be about 10 MeV.  
It is concluded that we suggested the possible $^{28}$Si-$^{12}$C molecular 
resonances and their connection with the low-lying 
superdefomation in $^{40}$Ca.

As mentioned above, the molecular states 
in the highly excited states in $^{28}$Si and $^{40}$C 
were discussed in the connection with the low-lying deformed states.
Here, we briefly report the possibility of molecular states 
in other $sd$-shell nuclei. 
With the same framework of AMD+GCM, we study the excited states of 
$^{36}$Ar and $^{28}$Mg.
In the highly excited states of $^{36}$Ar, we predict an eccentric cluster 
states with the 3-carbon-like linear structure. To our surprise, 
the main component of 
the 3-$^{12}$C linear-chain state is a $8p$-$12h$ configuration. It means that
the linear-chain state is explained also by the  many-particle many-hole state in
the picture of single-particle orbits. It is similar to the case of
the pear-like shape of the superdeformation of $^{40}$Ca, which is dominated
by the $8p$-$8h$ configuration. The linear-chain structure in $^{36}$Ar
is understood also by the two-body cluster picture. Considering that 
two of the three C clusters compose a $^{24}$Mg cluster, the linear-chain
is regarded as the excited $^{12}$C+$^{24}$Mg cluster states, where 
the C cluster contacts to the $^{24}$Mg from the longitudinal direction.
The rotational band of the linear structure is predicted in the
energy region $E_x\sim$30 MeV in the present calculations. 
Concerning $^{28}$Mg nucleus, various cluster channels may appear due to the 
valence neutrons and they interplay to each other.
A superdeformation with $2\hbar\omega$ configuration is theoretically 
obtained in the highly excited states of $^{28}$Mg. 
The results imply the possibility
of $^{14}$C-$^{14}$C and $^{18}$O-$^{10}$Be 
molecular states and $\alpha$-cluster states above the superdeformation. 
The coupling of the different channels, the fragmentation and the widths 
of the states should be investigated in detail.

\section{Summary}

Based on  microscopic calculations with the AMD method, we discussed
the excited states of $^{12}$Be, $^{28}$Si, $^{40}$Ca, $^{36}$Ar and
$^{28}$Mg while focusing on the cluster features.
Both the cluster aspect and mean-field aspect are essential in the systematic
study of these nuclei. These two natures coexist, compete, interplay or 
overlap to each other even in $sd$-shell nuclei as well as $p$-shell nuclei.
The largely deformed states are described by many-particle many-hole 
configurations, while the exotic shapes in the deformed states arise from 
the cluster structure.
We succeeded to connect the high-lying molecular states with the low-lying
deformed states. Namely,
the low-lying deformed states are regarded as lower members of the 
high-lying molecular states.
In other words, the molecular states arise due to the excitation of
inter-cluster motion and the higher molecular states start from the low-lying
deformed states.

Various molecular states were suggested in $sd$-shell nuclei.
To our impression, the carbon is the key cluster in $sd$-shell region
because the carbon clusters are often found in the present results. 
Needless to say, the $\alpha$ cluster is the most important cluster.
In addition to the $\alpha$ cluster we propose that the second key
cluster is carbon at least in $sd$-shell nuclei.

The coexistence of the cluster and mean-field natures
may become increasingly important in unstable nuclei. 
The further studies on unstable $sd$-shell nuceli, $^{20}$Ne and $^{32}$S 
were done by Kimura et al. with the extended version of 
AMD method(Ref.\cite{AMDrev}).

\end{document}